\documentstyle[pra,aps]{revtex}
\begin{document}
\draft 
\input epsf

\title{Probing Planckian physics: resonant production of particles\\
     during inflation and features in the primordial power spectrum}

\author{Daniel J.\ H.\ Chung\thanks{Electronic mail: djchung@umich.edu}}
\address{Randall Physics Laboratory,
     University of Michigan, Ann Arbor, Michigan \ 48109-1120}
     
\author{Edward W.\ Kolb\thanks{Electronic mail: rocky@rigoletto.fnal.gov}}
\address{NASA/Fermilab Astrophysics Center, Fermi
     National Accelerator Laboratory, Batavia, Illinois \  60510-0500,\\
     and Department of Astronomy and Astrophysics, Enrico Fermi Institute, \\
     The University of Chicago, Chicago, Illinois \ 60637-1433}

\author{Antonio Riotto\thanks{Electronic mail: riotto@nxth04.cern.ch}}
\address{Theory Division, CERN, CH-1211 Geneva 23, Switzerland}

\author{Igor I.\ Tkachev\thanks{Electronic mail: Igor.Tkachev@cern.ch}}
\address{Theory Division, CERN, CH-1211 Geneva 23, Switzerland, and\\
      Institute for Nuclear Research, Russian Academy of Sciences, Moscow 
      117312, Russia}

\date{September 1999}

\maketitle

\begin{abstract}
The phenomenon of resonant production of particles {\it after}
inflation has received much attention in the past few years.  In a new
application of resonant production of particles, we consider the
effect of a resonance {\em during} inflation.  We show that if the
inflaton is coupled to a massive particle, resonant production of the
particle during inflation modifies the evolution of the inflaton, and
may leave an imprint in the form of sharp features in the primordial
power spectrum.  Precision measurements of microwave background
anisotropies and large-scale structure surveys could be sensitive to
the features, and probe the spectrum of particles as massive as the
Planck scale.
\end{abstract}

\pacs{PACS: 98.80.Cq \hfill  FNAL-Pub-99/308-A;  CERN-TH/99-302}

\def\dslash{\partial{\!\!\!/}}
\def\simlt{\stackrel{<}{{}_\sim}}
\def\simgt{\stackrel{>}{{}_\sim}}
\def\identity{1 \hspace{-.085cm}{\rm l}}

\section{INTRODUCTION AND MOTIVATION}

Analysis of the statistics of temperature fluctuations of the cosmic
background radiation (CBR) and the distribution of galaxies and galaxy
clusters in large-scale structure (LSS) surveys yield the presently
observed power spectrum of scalar density perturbations.  The power
spectrum is a most useful and fundamental tool for understanding the
origin and development of the large-scale structure of the universe.

It is likely that the observed power spectrum had a primordial origin,
resulting from events that occurred in the early universe.  The
primordial power spectrum is not directly observable since it is
adulterated by a variety of cosmological and astrophysical processes
associated with the transition from a radiation-dominated to a
matter-dominated universe, with the decoupling of matter and
radiation, with the free streaming of dark matter, and so on.
Nevertheless, present observations of the power spectrum allow us to
infer the form of the primordial power spectrum.

To a first approximation, present observations of the power spectrum
are consistent with it having originated from a primordial power
spectrum that is featureless, i.e., a primordial spectrum of the form
$k^n$ where $k$ is the wavenumber and $n$ is the spectral index.
However, there are hints of features in the power spectrum from
analysis of the Automatic Plate Measurement (APM) survey \cite{apm},
from apparent regularity in the redshift distribution of galaxies seen
in deep pencil-beam surveys \cite{pencil}, cluster redshift surveys
\cite{clusters}, and galaxy surveys \cite{Century,LCRS}.

There are (at least) three possible responses to these ``hints.''  The
first response is that the apparent features are an artifact of small
data sets or data sets with systematic errors, and the effects will
eventually disappear with better or more complete observations.
Another possibility is that the features are real, and they reflect
the shape of the primordial power spectrum.  A third response is that
the primordial power spectrum is smooth, but astrophysical processing
of the primordial power spectrum produces the features.

A well known example of the last phenomenon is the ``Zel'dovich
peaks'' in the angular power spectrum of CBR temperature fluctuations.
The peaks are not believed to reflect features in the primordial
power spectrum, but presumably result from astrophysical processing,
in this case, acoustic oscillations of the baryon--photon fluid.
However, the features in the matter power spectrum
around 100$h^{-1}$\,Mpc are not well fit by acoustic oscillations
\cite{Danetal}.

The leading theory for the origin of the primordial power spectrum is
quantum fluctuations during inflation.  Simple models of inflation
produce a featureless, nearly exact power-law spectrum of primordial
perturbations.  Of course, an important qualifier in the previous
sentence is ``simple.''  A variety of inflation models that produce
features in the primordial spectrum have been studied.  These include
variants of extended inflation \cite{lasteinhardt} that produce rare
large voids \cite{roma}, models with multiple episodes of inflation
\cite{multiple}, and  phase transitions during the
inflationary phase \cite{alexi}.

In this paper we propose a mechanism to produce features in the
primordial power spectrum by a very simple variant of the simplest
inflation model.  Our basic idea is to couple the inflaton to a
massive particle, with resonant production of the particle during
inflation.  Resonant extraction of even a small fraction of the energy
of the inflaton field during inflation can alter the classical motion
of the inflaton and produce features in the primordial power spectrum.

The primordial power spectrum is most conveniently expressed in terms
of the amplitude of the density perturbation when it crosses the
Hubble radius after inflation, $\delta_H(k)$.  For reference, the
Harrison--Zel'dovich spectrum has $\delta_H(k)$ independent of $k$.  A
qualitative feel for our mechanism can be obtained by recalling the
simple approximation for the density perturbations produced during
inflation \cite{book},
\begin{equation}
\label{eq:simple}
\delta_H(k) \sim \frac{H^2}{5\pi\dot{\phi}} \ .
\end{equation}
Here, $H$ is the expansion rate of the universe and $\dot{\phi}$ is
the velocity of the inflaton field when the comoving wavenumber $k$
crossed the Hubble radius during inflation.  We find that resonant
extraction of inflaton field energy decreases $\dot{\phi}$, leading to
an increase in $\delta_H(k)$.  Because of the resonant nature of the
process, the produced feature is rather sharp, extending less than a
decade in wavenumber.

There are two reasons why our study may be important.  The first
reason is that features in the observed power spectrum have important
astrophysical implications.  Fitting the CBR Zel'dovich peaks is an
important part of cosmological parameter estimation
\cite{parameterestimation}.  Spurious peaks in the spectrum arising
from resonant particle production will complicate the extraction of
parameters (although joint parameter estimation from CBR and LSS
\cite{complementarity} may help alleviate the problem).

If baryons are a relatively large fraction of the matter density, the
effect of acoustic oscillations on the matter power spectrum
\cite{complementarity,acoustic} may be seen in accurate determinations
of the power spectrum in the next generation of large-scale structure
surveys such as the 2-degree Field project \cite{2df} and the Sloan
Digital Sky Survey (SDSS) \cite{sdss}, and may be part of the
parameter estimation program \cite{paramgal}.  Again, this program is
complicated if the primordial perturbation spectrum does not have a
smooth power law form \cite{wang}.  Finally, the program of
reconstruction of the inflaton potential from CBR and large-scale
structure determinations of the power spectrum
\cite{recon1,recon2,recon3} may be compromised if there are sharp
features.

The second reason we believe our study is relevant is that peaks in
the primordial power spectrum may serve as a probe of energy scales as
large as the Planck scale.  We will discuss the particle physics
motivation for the coupling and masses of the particles in a later
section.  But to see the possibility of probing Planck-scale physics,
simply assume a chaotic inflation model with a quadratic
potential. The inflaton mass in this model is around $10^{13}$\,GeV,
which is also approximately the expansion rate of the universe during
inflation.  During inflation the value of the inflaton field is a few
times the Planck mass, $m_{Pl}$.  Since the energy density during the
inflationary stage is of the order of $m_\phi^2m_{Pl}^2\ll m_{Pl}^4$,
in the spirit of effective field theories we may regard the inflaton
potential as the low-energy limit of some fundamental Planck-scale
theory. Furthermore, values of the the inflaton field $\phi\sim
m_{Pl}$ do not necessarily lead to a breakdown of the field theory
expansion; for example, moduli fields in superstring theories whose
vacuum expectation values parameterize different string vacua usually
have field values of the order of the Planck scale.

The inflaton may be coupled to other degrees of freedom with mass a
few times the Planck mass, but these degrees of freedom are usually
integrated out and do not appear in the low-energy action.  Of
relevance for our considerations is the possibility that the inflaton
field couples to a particle of mass a few times the Planck mass.
Normally, the massive particle would be integrated out and not affect
the low-energy theory.  In the inflaton models we study, observable
length scales exit the Hubble radius during inflation when the value
of the inflaton field is a few times the Planck mass.  If the inflaton
couples to a fermion of mass $m$ (assumed to be of order of $m_{Pl}$)
with Yukawa coupling $\lambda$, then the $\phi$-dependent mass of the
fermion is $M(\phi) = m - \lambda \phi$.\footnote{The important
feature here is that the $\phi$-dependent mass vanish during
inflation.  This can be accomplished by choosing a negative sign for
the Yukawa term, or by taking $\phi$ to be negative.  In this paper we
will choose positive value of $\phi$ and a negative Yukawa term.  An
important notational convention is the distinction between $m$ and
$M(\phi)=m-\lambda\phi$.}  The $\phi$ ground state is $\phi=0$, and in
the ground state the fermion is supermassive and not part of the
effective low-energy theory.  But when $\phi \simeq m/\lambda$, the
effective mass of the fermion vanishes, and it cannot be integrated
out.  When $\phi \simeq \phi_* = m/\lambda$ and the fermions are
light, and they can be efficiently produced \cite{gprt}.  Since $\phi
\simgt m_{Pl}$ during inflation, it can act as a probe of Planckian
physics.

Ours is not the first study of parametric creation of fermions.  Pure
gravitational creation was considered in Refs.\
\cite{mmf,KT99}. Creation by an oscillating background field in
Minkowski space was considered in Refs.\ \cite{min1,min2}.  Resonant
production of {\it massless} fermions {\it after} inflation (i.e.,
during preheating) in a $\lambda\phi^4$ inflation model was studied by
Greene and Kofman \cite{greenekofman}. Non-thermal productions of
gravitinos have been considered in Ref.\ \cite{gra}. Finally, resonant
production of {\it massive} fermions during preheating {\it after}
inflation in an $m_\phi^2\phi^2$ inflation model was considered by
Giudice, Peloso, Riotto, and Tkachev \cite{gprt}.  We draw heavily
from the results in these important papers.  The new twist we present
is a change in the venue for particle production: we assume the
resonance occurs {\em during} inflation.  We believe this may be the
most easily tested possibility because it leads to directly observable
effects in the power spectrum.

In the next section we will provide more details about the couplings
in the model we study and discuss the basics of resonant particle
creation during inflation.  In Sec.\ III we consider resonant
production where back reactions are not important.  We make some
analytic approximations leading to an estimate of the effect on the
density perturbation spectrum.  In Sec.\ IV we present the results of
a numerical calculation of the perturbation spectrum including the
effect of resonant particle creation in the regime where back
reactions are important.  In Sec.\ V we discuss the prospect of
detection of the type of features we produce.  Finally we present a
brief summary in the concluding section.

\section{HEAVY PARTICLE PRODUCTION DURING INFLATION}

Resonant particle production can occur for a range of choices of
inflation models as well as different heavy particle masses and
couplings of the massive state to the inflaton.  We will develop
analytic techniques and present numerical results for one model, and
mention the generalization of our results where appropriate.

We will assume that the inflaton potential during inflation is given
by a simple quadratic potential of the form
\begin{equation}
\label{eq:inflatonpotential}
V(\phi) = \frac{1}{2}m_\phi^2\phi^2 \ .
\end{equation}
The value of the inflaton mass can be fixed by normalization of the
perturbation spectrum, with result $m_\phi = 10^{13}$\,GeV.  Inflation
occurs for Planckian values of the inflaton field, $\phi \simgt
m_{Pl}$, where $m_{Pl}$ is the Planck mass.  Inflation ends when $\phi
\sim 0.2m_{PL}$, and scales relevant for CBR fluctuations and
large-scale structure studies cross the Hubble radius during inflation
when $\phi$ is in the range of approximately $2m_{Pl}$ to $3m_{Pl}$.

Now we will couple a fermion $\psi$ of mass $M$ to the inflaton
by a Yukawa term of the form\footnote{Here we treat a Dirac fermion,
but there is no qualitative difference if one chooses Majorana
fermions.}
\begin{equation}
\label{eq:yukawa}
{\cal L}_Y = - \lambda \phi \bar{\psi} \psi \ .
\end{equation}
If $\phi$ is nonzero, then the effective mass of the fermion
is\footnote{Notice that radiative corrections from the heavy fermion
may spoil the flatness of the inflaton potential. As usual, one can
invoke supersymmetry to preserve the flatness. If this is the case
and the starting superpotential is $2W= m_\phi\phi^2 +
\left(m-\lambda\phi\right)\psi^2$, it is easy to show that radiative
corrections are negligible, with the one-loop effective potential
$32\pi^2V_1\simeq \lambda^2 m_\phi^2\phi^2
\log(\lambda^2\phi^2/\Lambda^2)$, where $\Lambda$ is the
renormalization scale.}
\begin{equation}
\label{eq:effectivemass}
M(\phi) = m - \lambda \phi \ .
\end{equation}
These choices lead to a critical value of the inflaton field, which we
will denote as $\phi_*$, where the effective mass of the fermion
vanishes:
\begin{equation}
\label{eq:phistar}
\phi_* = m / \lambda \ .
\end{equation}
Resonant fermion production will occur in a narrow interval around
$\phi = \phi_*$ \cite{gprt}.

Although we chose a fermion for the massive state, one might imagine
that the massive state is a boson.  Recall that the requirement for
particle production is that the effective mass term goes through zero
at $\phi_*$.  If the effective mass term goes through zero, then one
might expect that it is negative either above or below $\phi_*$.
Since the sign of a fermion mass term is irrelevant, a negative mass
term is not a problem.  If the massive state is a boson, then the
quadratic nature of the mass term implies one has to worry about
tachyonic modes.  So if the massive state is a boson, then the
effective mass term should remain positive both above and below
$\phi_*$.  This can be done, for instance, by taking a boson $\sigma$
with potential $V(\sigma) = \sigma^2 \left(m_\sigma - g\phi
\right)^2+\lambda_\sigma\sigma^4$, where $m_\sigma > 0$. The vacuum
expectation value of the field $\sigma$ vanishes. The effective mass
term of the boson, $M_\sigma^2 = 2\left(m_\sigma - g\phi \right)^2$,
remains positive definite and vanishes at $\phi_*=m_\sigma/g$.

Now consider fermion production.  In an expanding universe, the
fermion field $\psi$ satisfies the Dirac equation in conformal time
$\eta$ ($a d\eta = dt$),
\begin{equation}
\label{eq:dirac}
\left( \frac{i}{a} \dslash + i \frac{3}{2}H\gamma^0-M \right)\psi 
= 0 \ ,
\end{equation}
where $a$ is the scale factor, $\gamma^\mu$ are the flat-space gamma
matrices, the spatial derivatives are with regard to comoving
coordinates, and the Hubble expansion rate is $H=\dot{a}/a= a'/a^2$
(here and throughout the paper a prime superscript implies $d/d\eta$).
This equation can be reduced to a more familiar form of the Dirac
equation by defining a new field $\chi = a^{-3/2}\psi$.  In terms of
the new field $\chi$, the Dirac equation in an expanding universe
becomes
\begin{equation}
\left[ i \dslash -a(\eta)M(\eta) \right] \chi = 0 \ .
\end{equation}
Of course $M$ depends on $\eta$ through the dependence of $\phi$
on $\eta$.

The field $\chi$ can be expanded in terms of Fourier modes of the form
\begin{equation}
\label{eq:diractwo}
\chi(x) = \int \frac{d^3\!k}{(2\pi)^{3/2}}\ e^{-i\vec{k}\cdot\vec{x}}
\, \sum_{r=\pm 1} \left[ u_r(k,\eta)a_r(k) +
v_r(k,\eta)b^\dagger_r(-k)\right] \ ,
\end{equation}
where the summation is over spin and $v_r(k) \equiv C\bar{u}^T_r(-k)$.
The canonical anticommutation relations imposed upon the creation and
annihilation operators may be used to normalize the spinors $u$ and
$v$.

Defining $u_r \equiv \left[u_+(\eta)\psi_r(k), r u_
-(\eta)\psi_r(k)\right]^T$ and $v_r \equiv \left[r v_+(\eta)\psi_r(k),
v_-(\eta)\psi_r(k)\right]^T$, where $\psi_r(k)$ are the two-component
eigenvectors of the helicity operators, and using a representation
where $\gamma^0 = {\rm diag}(\identity,-\identity)$, Eq.\
(\ref{eq:diractwo}) can be written as two uncoupled second-order
differential equations for $u_+$ and $u_-$:
\begin{equation}
\label{eq:waveequation}
u_\pm'' + \left[ \omega_k^2 \pm i (aM)' \right] u_\pm = 0 \ ,
\end{equation}
where, $\omega_k^2(\eta) = k^2 + M^2(\eta)a^2(\eta)$ and $M(\eta) = m
- \lambda\phi(\eta)$. Notice that the normalization conditions
$u_r^\dagger u_s=v_r^\dagger v_s=2\delta_{rs}$ and $u_r^\dagger v_s=0$
are preserved during the evolution. Furthermore, sometimes it is
useful to remember that $v_{+}$ has the same evolution as $-u_{-}^*$
and $v_{-}$ has the same evolution as $u_{+}^*$.

In order to calculate the number density, we must first diagonalize
the Hamiltonian.  In the basis of Eq.\ (\ref{eq:diractwo}) the
Hamiltonian is
\begin{equation}
H(\eta) = \int d^3\!k \sum_r \left\{
E_k(\eta) \left[ a_r^\dagger(k)a_r(k) - b_r(k)b_r^\dagger(k) \right] +
F_k(\eta)b_r(-k)a_r(k) 
+ F_k^*(\eta)a_r^\dagger(k)b_r^\dagger(-k)\right\} \ ,
\end{equation}
where the equations of motion can be used to express $E_k$ and $F_k$
in terms of $u_+$ and $u_-$:\footnote{Here we choose the momentum $k$
along the third axis and use the representation in which $\gamma^3 =
\left( \begin{array}{cc} 0 & \identity \\ -\identity & 0 \end{array}
\right)$.}
\begin{eqnarray}
E_k & = & k {\rm Re}(u_+^*u_-) + 
aM\left(1-\left|u_+\right|^2\right) \ , \nonumber \\
F_k & = & \frac{k}{2}(u_+^2-u_-^2) +aM u_+u_- \ .
\end{eqnarray}

In order to calculate particle production one wants to write the
Hamiltonian in terms of creation and annihilation operators that are
diagonal. To do this one defines a new set of creation and
annihilation operators, $\hat{a}$ and $\hat{b}^\dagger$, related to
the original creation and annihilation operators $a$ and $b^\dagger$
through the (time-dependent) Bogolyubov coefficients $\alpha_k$ and
$\beta_k$,
\begin{eqnarray}
\label{sy}
\hat{a}(k) & = & \alpha_k(\eta) a(k) + \beta_k(\eta)b^\dagger(-k) \ ,
                \nonumber \\
\hat{b}^\dagger(k) 
       & = & -\beta^*_k(\eta) a(k) + \alpha^*_k(\eta)b^\dagger(-k) \ . 
\end{eqnarray}
The Bogolyubov coefficients will be chosen to diagonalize the
Hamiltonian.  Using the fact that the canonical commutation relations
imply $|\alpha_k|^2 + |\beta_k|^2=1$, the choice
\begin{equation}
\label{choice}
\frac{\alpha_k}{\beta_k} = \frac{E_k+\omega_k}{F^*_k} \ , \quad
\left| \beta_k \right|^2 = \frac{|F_k|^2}{2\omega_k(\omega_k+E_k)} \ ,
\end{equation}
results in a diagonal Hamiltonian,
\begin{equation}
H(\eta) = \int d^3\!k \sum_r \omega_k(\eta) \left[
\hat{a}_r^\dagger(k)\hat{a}_r(k) + \hat{b}^\dagger_r(k)\hat{b}_r(k) 
\right] \ . 
\end{equation}

We define the initial vacuum state $| 0 \rangle$ such that ${a} |
0 \rangle = {b} | 0 \rangle = 0$.
The initial conditions corresponding to the no-particle state are
\begin{equation}
u_\pm(0)=  \sqrt{\frac{\omega_k\mp M a}{\omega_k}} \ ; \qquad
u'_\pm(0) = iku_\mp(0)\mp iaM u_\pm(0) \ . 
\end{equation}

The (quasi) particle number operator can be defined as ${\cal N} =
\hat{a}_r^\dagger(k)\hat{a}_r(k)$, and the particle number density $n$
is (including the two degrees of freedom from the spin)
\begin{equation}
\label{number}
n(\eta) = \langle 0 \left | {\cal N}/V \right | 0 \rangle =
\frac{1}{\pi^2a^3(\eta)} \int_0^\infty dk \, k^2 
\left| \beta_k \right|^2.
\end{equation}
An  equal amount of antiparticles is produced.

Including the coupling of the fermion to the inflaton, the inflaton
equation of motion in the Hartree approximation is
\begin{equation}
\label{eq:eom}
\ddot{\phi} + 3 H\dot{\phi} + \frac{dV}{d\phi} 
- N\lambda \langle \bar{\psi}\psi \rangle = 0 \ ,
\end{equation}
where we have generalized to the case of $N$ fermions of bare mass $m$
coupled to the inflaton.

The product $\langle \bar{\psi} \psi \rangle$ can be expressed as
momentum integration of the mode functions using the field
decomposition, Eq.\ (\ref{eq:diractwo}). A straightforward averaging
leads to ultraviolet divergences, therefore, the quantity $\langle
\bar{\psi} \psi \rangle$ must be regularized. As in the case of
Minkowski spacetime, the regularization amounts to normal ordering or,
equivalently, to the subtraction of zero-point vacuum fluctuations. To
obtain a finite result, it is necessary to express the operator $
\bar{\psi} \psi $ in normal-ordered form and subtract the part due to
vacuum fluctuations.  The normal-ordered $ \bar{\psi} \psi $ operator
has the form
\begin{equation}
N_\eta\left(\bar{\psi} \psi \right) \equiv  \bar{\psi} \psi - 
\langle 0_\eta | \bar{\psi} \psi |0_\eta\rangle  \,  ,
\end{equation}
where the vacuum $ |0_\eta\rangle $ is defined by 
\begin{equation}
\hat a  
|0_\eta\rangle
=\hat b |0_\eta\rangle =0 \, . 
\label{vacc}
\end{equation}

The vacuum averaging in Eq.\ (\ref{eq:eom}) is defined as averaging
with respect to the original vacuum state (we remind the reader that
we are working in the Heisenberg representation)
\begin{equation}
\langle \bar{\psi} \psi \rangle \equiv \langle 0| N_\eta
\left( \bar{\psi} \psi \right) |0\rangle \,.
\end{equation}
Using the operator redefinitions given in Eq.\ (\ref{sy}), it is
straightforward to show that
\begin{equation}
\langle  \bar{\psi} \psi \rangle = \frac{4}{(2\pi a)^3}\int d^3k
\left[\left(|u_+|^2-1\right)\left|\beta_k\right|^2+{\rm Re}\left(
\alpha_k\beta_k^* u_{+}^* u_{-}^*\right)
\right].
\end{equation}
Making use of the formulae in Eq.\ (\ref{choice}), we arrive at the
following expression for the regularized average of $\bar{\psi} \psi$:
\begin{equation}
\label{eq:pp}
\langle  \bar{\psi} \psi \rangle = \frac{2}{(2\pi a)^3}\int d^3k 
\left(|u_+|^2 + \frac{Ma}{\omega_k} -1 \right) \,.
\end{equation}
In the case of Majorana spinors the numerical factor is half that of
Eq.\ (\ref{eq:pp}).

We will use Eq.\ (\ref{eq:pp}) when integrating the equation of motion
[given in Eq.\ (\ref{eq:eom})] for the inflaton field. We have also
consistently included the contribution from $\langle \bar{\psi} \psi
\rangle$ in the equation of state when integrating Einstein's
equations for the scale factor. We integrated the system of all
equations in the conformal frame with Eq.\ (\ref{eq:waveequation}) the
equation of motion for the mode functions.  For the case of an
inflaton potential given by Eq.\ (\ref{eq:inflatonpotential}), Eq.\
(\ref{eq:eom}) takes the form
\begin{equation}
{\varphi}'' + \left(m_\phi^2a^2 - \frac{a''}{a} \right)\varphi 
- a^3 N\lambda \langle \bar{\psi} \psi \rangle = 0 \, ,
\end{equation}
where $\varphi \equiv a \phi $.

At this point it is useful to discuss the unknown parameters of the
calculation.  First, we do not know the inflaton potential ($V$).  We
simply assume it is a quadratic large-field model for the purposes of
detailed calculations and return to this issue in a later section.  We
do not know the mass of the fermion ($m$) and the Yukawa coupling of
the fermion to the inflaton ($\lambda$).  We will assume that $m$ and
$\lambda$ are such that $\phi_*$, defined in Eq.\ (\ref{eq:phistar}),
falls in the range of $\phi$ that corresponds to the epoch of
inflation when astrophysically accessible scales are crossing the
Hubble radius. In the large-field inflation model we study, this
corresponds to $m/\lambda$ a few times the Planck mass.  Particle
production is most efficient when $\lambda$ is large, so $m$ also must
be large to have $\phi_*$ in the relevant region.  Finally, we do not
know the number of fermion fields that couple to the inflaton ($N$).
This number is potentially very large because there are often fermion
representations of large dimension in superunified theories.  So with
the assumption of the large-field quadratic inflation model, the
parameter space is spanned by $\{m,\,\lambda,\,N\}$, with the ratio
$m/\lambda$ reasonably well constrained.

\section{ANALYTIC APPROXIMATIONS AND NUMERICAL RESULTS WHEN BACK 
REACTIONS ARE UNIMPORTANT}

If only a small fraction of inflaton energy is extracted in resonant
production of particles, then it is possible to find reasonably
accurate analytic approximations for the features produced in the
primordial perturbation spectrum.

There are three steps in obtaining analytic approximations for the
modification to the perturbation spectrum due to resonant particle
production during inflation.  The first step is to determine the
efficiency of resonant fermion production.  The next step is to
estimate the effect of fermion production on $\dot{\phi}$.  Finally,
it is necessary to convert a change in $\dot{\phi}$ to the
modification of the perturbation spectrum.

\subsection{Fermion Production}

First consider the question of particle production.  Start with the
wave equation, Eq.\ (\ref{eq:waveequation}):
\begin{equation}
u_\pm'' + (k^2 + M^2a^2)u_\pm \pm  i (a M)'u_\pm = 0.
\end{equation}
Now define the (conformal) time $\eta_*$ to be the time when
$M(\eta_*)=0$.  Recalling that $M=m-\lambda \phi$, in
the vicinity of $\eta=\eta_*$ we may expand $M(\eta)$ as
\begin{equation}
\label{eq:expand}
M(\eta) = \lambda \phi_*'(\eta-\eta_*) +\cdots \ ,
\end{equation}
where $\phi_*' \equiv \phi'(\eta_*)$.  Now let's assume that resonant
production happens in a very narrow region of conformal time and the
scale factor does not change much during particle production (i.e.,
$a'=0$).  Choosing $a_* \equiv a(\eta_*) = 1$, then around $\eta_*$
the wave equation is
\begin{equation}
u_\pm''+ \left[ k^2 + \lambda^2{\phi_*'}^2 (\eta-\eta_*)^2\right]u_\pm 
\mp i \lambda \phi_*'u_\pm = 0 \ .
\end{equation}
Now defining new variables $p = k/\sqrt{\lambda\phi_*'}$, $\tau =
\sqrt{\lambda\phi_*'} (\eta-\eta_*)$, and denoting $d/d\tau$ by a dot,
the wave equation becomes
\begin{equation}
\ddot{u}_\pm + \left( p^2 \mp i + \tau^2 \right) u_\pm = 0 \ .
\end{equation}
The solutions to this equation are the parabolic cylinder functions.
Choosing the boundary conditions associated with the no-particle
initial state, one can estimate the phase-space density of the created
particles to be (cf.\ the Bosonic case discussed in Refs.\
\cite{birrellanddavies,KLS97,danthesis})
\begin{equation}
\label{eq:analytic}
\left| \beta_k \right|^2 \simeq
\exp \left( -\frac{\pi k^2}{\lambda|\phi_*'|} \right)  \ .
\end{equation}

The argument of the exponential appearing in Eq.\ (\ref{eq:analytic})
can be expressed in terms of a physical momentum, $k_p$, the time
derivative of the inflaton field, and the expansion rate at resonance,
as
\begin{equation}
\frac{\pi k^2}{\lambda \phi_*'} = \frac{\pi}{\lambda}\,
\frac{H_*^2}{|\dot{\phi}_*|} \, \left(\frac{k_p}{H_*}\right)^2 \ .
\end{equation}
Here $H_*$ is the Hubble expansion rate and $\dot{\phi}_*$ is the 
unperturbed velocity of the inflaton field at $\eta=\eta_*$.  

\begin{figure}[t]
\centering \leavevmode\epsfxsize=350pt \epsfbox{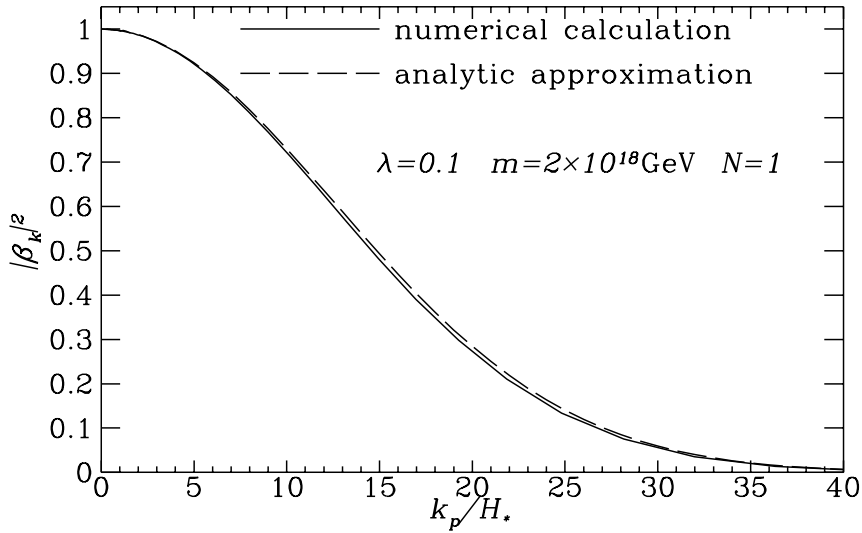}
\caption[fig1]{\label{spectrum} The spectrum of produced fermions
around $\eta=\eta_*$ as a function of physical momentum $k_p$.  Also
shown is the analytic approximation.}
\end{figure}

If the resonance occurs at $\phi_* \simeq 2m_{Pl}$, then $H_*\simeq 4
m_\phi$ and $|\dot{\phi}| \simeq 0.16 m_{Pl}m_\phi$. The crucial fact
is that in this model the ratio $H_*^2/|\dot{\phi}_*|$ is very small,
about $10^{-4}$.  This allows the phase-space density to be large in
regions where the physical momentum is larger than the inverse of the
Hubble radius.  The smallness of this ratio should obtain in all
slow-roll inflation models.

Fig.\ \ref{spectrum} shows an example of the comparison between the
analytic formula of Eq.\ (\ref{eq:analytic}) and a numerical solution
of the equations.  The momentum is in units of $k_p/H_*$, where $k_p$ is
the physical momentum at the time $\eta_*$.

There are two important results one may learn from Fig.\
\ref{spectrum}.  The first is that the analytic approximation is quite
accurate.  The second important fact is that, as expected from the
discussion above, the phase-space density is large for $k$ greater
than $H_*$.\footnote{The contribution per logarithmic interval in $k$
is proportional to $k^3|\beta_k|^2$, so the peak contribution to the
number density of created particles occurs for $k_p\gg H_*$.}  This
means that the effect is dominated by sub-Hubble-radius wavelength
modes.

The physical number density of each Dirac fermion species (particles
plus antiparticles for a total of four degrees of freedom) produced at
$\eta_*$ is given by
\begin{equation}
\label{eq:nstar}
n(\eta_*) = \frac{2}{\pi^2}\int_0^\infty dk_p \, k_p^2 \, |\beta_k|^2 = 
\frac{\lambda^{3/2}}{2\pi^3}
\left(\frac{|\dot{\phi}_*|}{H_*^2}\right)^{3/2}H_*^3 \ .
\end{equation}
For $\eta \geq \eta_*$, the number density of fermions simply
decreases as $a^{-3}$.  For $\eta>\eta_*$ the fermion mass is
approximately $m$, so knowing the number density at $\eta=\eta_*$
allows us to estimate the contribution of the produced fermions for
$\eta>\eta_*$.

\subsection{Inflaton Velocity}

Now we turn to the issue of how resonant production of the fermions
modifies $\dot{\phi}$.  Let's start with the equation of motion for
the inflaton field, Eq.\ (\ref{eq:eom}).  Here we are interested in
change in the velocity of the inflaton field caused by particle
creation.  Consider a small region around the time of particle
creation, $t=t_*$.  In this small region $H=H_*$ and
$dV/d\phi=(dV/d\phi)_*$ are approximately constant. 

We may also approximate $\langle \bar{\psi}\psi \rangle$ by assuming
the number density of fermions is zero before $t_*$, equal to $n_*
\equiv n(\eta_*)$ [given by Eq.\ (\ref{eq:nstar})] at $t=t_*$, and
decreases as $a^{-3}$ thereafter:
\begin{equation}
\label{eq:matter}
\langle \bar{\psi}\psi \rangle = 
n_* \, \theta(t-t_*) \, \left( \frac{a_*}{a} \right)^3
\simeq n_* \, \theta(t-t_*) \, \exp[-3H_*(t-t_*)] \ . 
\end{equation}
Here for simplicity we assume an exponential growth of the scale
factor during inflation. 

The fact that $\langle \bar{\psi}\psi \rangle$ can be approximated by
$n_*$ may be understood as follows. The mode functions have an
adiabatic evolution before and after the time $\eta_*$ when
$M(\eta_*)=0$. The nonadiabatic changes of the mode functions occur
only in the vicinity of $\eta_*$. Therefore we expect the mode
function
\begin{equation}
u_{+}(\eta<\eta_*,k)=\sqrt{
1-\frac{Ma}{\omega_k}}\; \exp\left(-i\int^\eta \omega_k\:d\eta\right)
\end{equation}
to be the adiabatic solution of Eq.\ (\ref{eq:waveequation}) before the
particle production at time $\eta_*$. After the production time, the
mode function $u_{+}$ is again an adiabatic solution, with the
difference that it picks up a negative frequency. It may be expressed
in the form
\begin{equation} 
u_{+}(\eta>\eta_*,k)=\alpha_k \sqrt{
1-\frac{Ma}{\omega_k}}\; \exp\left(-i\int^\eta \omega_k\:d\eta\right) 
-\beta_k\sqrt{1+\frac{Ma}{\omega_k}}\;  
\exp\left(+i\int^\eta \omega_k\:d\eta\right) \ .
\end{equation}
Inserting this expression into Eq.\ (\ref{eq:pp}) and making use of
the property $\left|\alpha_k\right|^2+\left|\beta_k\right|^2=1$ , we
obtain
\begin{equation}
\label{ap}
\langle  \bar{\psi} \psi \rangle = \frac{4}{(2\pi a)^3}\int d^3k
\left\{ \frac{Ma}{\omega_k} \left|\beta_k\right|^2
-\frac{k}{\omega_k}{\rm Re}\left[\alpha_k\beta_k^*
\exp\left(-2i\int^\eta 
\omega_k\:d\eta\right)\right] \right\} \ .
\end{equation}
The Bogolyubov coefficient $\left|\beta_k\right|^2$ may be well
approximated by unity for $k\simlt
k_*=\sqrt{\lambda\left|\phi^\prime_*\right|}$ and zero
otherwise. Furthermore, at $\eta>\eta_*$ it is safe to neglect the
inflaton field dependent part in the mass $M$ of the fermion $\psi$
and we have $M\simeq m\gg k_*$.  Under these circumstances, 
Eq.\ (\ref{ap}) reduces to
\begin{equation}
\langle  \bar{\psi} \psi \rangle = \frac{4}{(2\pi a)^3}\int d^3k
\left(\frac{Ma}{\omega_k} \left|\beta_k\right|^2\right)\simeq
\frac{4}{(2\pi a)^3}\int d^3k \left|\beta_k\right|^2\equiv n_*.
\end{equation}
This proves that $\langle \bar{\psi}\psi \rangle$ can be
approximated by $n_*$.

With this approximation for $\langle \bar{\psi}\psi \rangle$, the
first-order differential equation for $\dot{\phi}$ [ Eq.\
(\ref{eq:eom})] is
\begin{equation}
\frac{d\dot{\phi}}{dt} + 3H_* \dot{\phi} + (dV/d\phi)_* - 
N\lambda n_* \theta(t-t_*)\, \exp[-3H_*(t-t_*)] = 0 \ .
\end{equation}
The solution of this equation for $t>t_*$ is 
\begin{equation}
\dot{\phi}(t>t_*) = \dot{\phi}_*\exp[-3H_*(t-t_*)] 
- \frac{(dV/d\phi)_*}{3H_*}\left\{ 1 - \exp[-3H_*(t-t_*)] \right\}
+ N \lambda n_* (t-t_*) \, \exp[-3H_*(t-t_*)] \ .
\end{equation}
We may easily calculate the change in $\dot{\phi}$ due to particle
creation:
\begin{equation}
\label{eq:deltaphidot}
\Delta\dot{\phi}(t>t_*)= \dot{\phi}(t>t_*) - 
\left[\dot{\phi}(t>t_*)\right]_{\lambda=0} =
N \lambda n_* (t-t_*) \, \exp[-3H_*(t-t_*)] \ .
\end{equation}

\begin{figure}[t]
\centering \leavevmode\epsfxsize=350pt \epsfbox{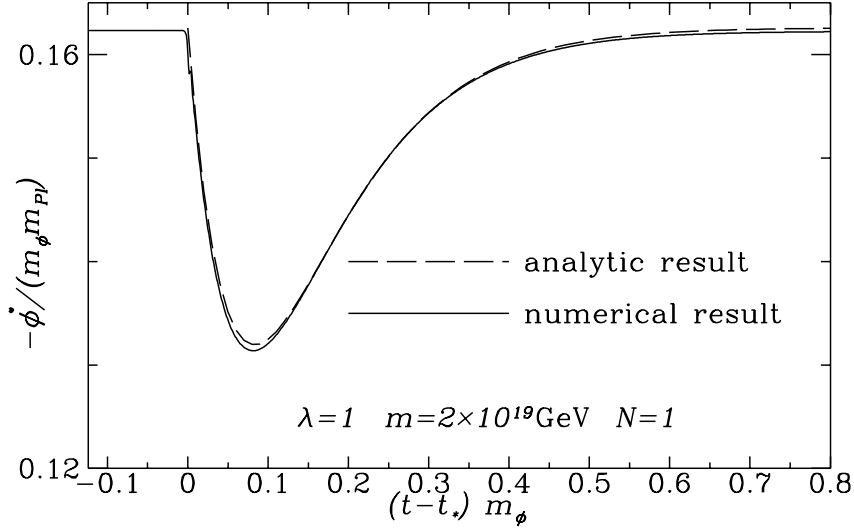}
\caption[fig2]{\label{phi_dot_1_0} The numerical result for the
evolution of $\dot{\phi}$, including the effect of resonant particle
production, is shown by the solid curve.  Also shown by the dashed
curve (nearly indistinguishable from the numerical result) is the
analytic expression of Eq.\ (\ref{eq:deltaphidot}). }
\end{figure}

The maximum size of the feature can also be
estimated. Equation (\ref{eq:deltaphidot}) has a maximum at
$t=(3H_*)^{-1}$ of
\begin{equation}
\label{eq:estimatemaxphidot}
\frac{{\rm max}[\Delta\dot{\phi}]}{m_{Pl}m_\phi} = 
\frac{\Delta\dot{\phi}(t=1/3H_*)}{m_{Pl}m_\phi} = 
\frac{N\lambda^{5/2}}{6e\pi^3}\, 
\frac{|\dot{\phi}_*|^{3/2}}{H_*m_{Pl}m_\phi} \ ,
\end{equation}
where we have used Eq.\ (\ref{eq:nstar}) for $n_*$.

In Fig.\ \ref{phi_dot_1_0} we present the result of a numerical
calculation.  Note that ${\rm max}[\Delta\dot{\phi}]$ is about
$0.03m_{Pl}m_\phi$.  For the model of Fig.\ \ref{phi_dot_1_0},
$H_*\simeq 4m_\phi$ and $\dot{\phi}_* \simeq 0.16 m_{Pl}m_\phi$, and
from Eq.\ (\ref{eq:estimatemaxphidot}) we predict ${\rm
max}[\Delta\dot{\phi}] \simeq 0.03m_{Pl}m_\phi$, in excellent
agreement with the numerical results.  Not just the depth of the
feature, but the entire shape of the feature is well approximated
by Eq.\ (\ref{eq:deltaphidot}).

We have assumed that the fermion is stable, or at least has a lifetime
long compared to the time it takes to form the feature, $t-t_* \simeq
0.2m_\phi^{-1}$.  This is not unreasonable at all. Indeed, in spite of
the fact that the bare mass of the fermion is much larger than the
inflaton mass, the fermions are produced when they are
massless. Suppose we parameterize the decay rate of the fermion field
by $\Gamma=\alpha M$, where $\alpha$ is a perturbative constant
smaller then unity. Using the expansion of Eq.\ (\ref{eq:expand}) and
imposing the condition $\Gamma\simgt H$, we estimate that the fermions
decay at time $t_D$ such that $t_D-t_*\simeq \left (\lambda\alpha
\phi_*\right)^{-1}$. The fermionic decays take place after the
formation of the features in the spectrum if $t_D-t_*\simgt
0.2m_\phi^{-1}$. This amounts to requiring $\alpha\simlt (5/\lambda)
(m_\phi/\phi_*)$, which is not very restrictive. Another option is
that the perturbative decay rate $\Gamma$ identically vanishes for
kinematical reasons at the inflationary stage. This will happen if the
particles coupled to the fermion $\psi$ are heavier than the mass of
the $\psi$ during inflation.

\begin{figure}[t]
\centering \leavevmode\epsfxsize=350pt \epsfbox{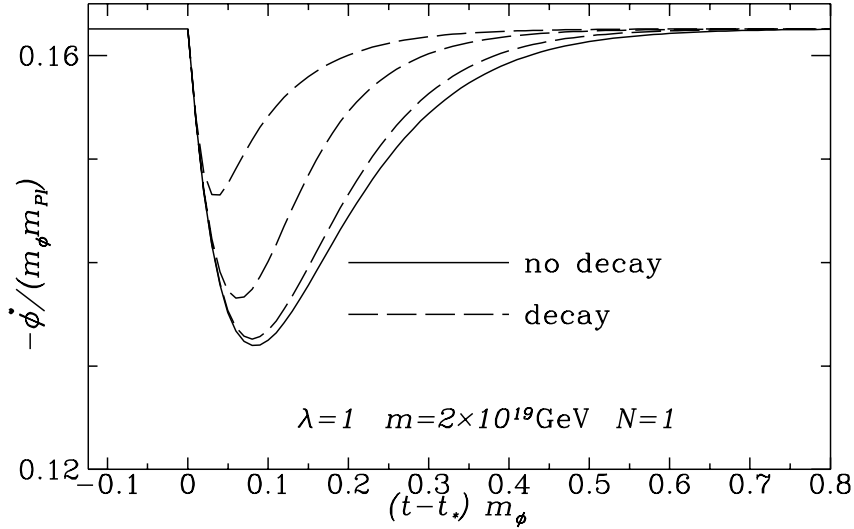}
\caption[fig3]{\label{decay} Analytic solutions for $\dot{\phi}$ assuming
no decay of the fermion (the solid curve) and with decay parameters
$\sqrt{\alpha\lambda\dot{\phi}_*}/m_\phi = 10^3$, $10^2$, and $10^1$.
The feature disappears as $\sqrt{\alpha\lambda\dot{\phi}_*}/m_\phi
\rightarrow \infty$, and as the quantity approaches zero, we recover
the result of no decay.}
\end{figure}

In the opposite extreme, $\Gamma\ll H_*$, we may assume that the
fermions decay very soon after they are produced.  In this case, the
factor of $\exp[-3H_*(t-t_*)]$ in Eq.\ (\ref{eq:matter}) would be
modified.  Again using the fact that around the resonance $M\simeq
\lambda
\dot{\phi}_*(t-t_*)$, the factor becomes
$\exp[-3H_*(t-t_*)]\exp[-\alpha\lambda\dot{\phi}_*(t-t_*)^2]$.  One can
follow through and derive a similar analytic approximation to
$\Delta\dot{\phi}$ [cf.\ Eq.\ (\ref{eq:deltaphidot})],
\begin{equation}
\label{eq:phidotdecay}
\Delta\dot{\phi} = N\lambda n_* \exp[-3H_*(t-t_*)] \,
\frac{1}{\sqrt{\alpha\lambda\dot{\phi}_*}} \,
\frac{\sqrt{\pi}}{2} \, 
{\rm erf}\left[ \sqrt{\alpha\lambda\dot{\phi}_*}(t-t_*)\right] \ .
\end{equation}
Of course in the limit that the particle is stable, $\alpha
\rightarrow 0$, the error function may be expanded to recover Eq.\
(\ref{eq:deltaphidot}).  The decay of the massive state results in a
feature that is narrower and less pronounced than if the massive
particle energy density decreases simply as matter (see Fig.\
\ref{decay}).  The feature in $\dot{\phi}$ will map onto the feature
in $\delta_H$ as discussed in the next section.

We do not know the relevant value of $\alpha$, but if we choose a
random value, say $\alpha = (137)^{-1}$, then
$\sqrt{\alpha\lambda\dot{\phi}_*} = 34$ for $\lambda=1.$ This would
make a few percent change in the amplitude of the feature and make it
a bit narrower (see Fig.\ \ref{decay}).

\subsection{Effect on the Perturbation Spectrum}

The simple expression for the perturbation amplitude in terms of $H$
and $\dot{\phi}$ given in the introduction, Eq.\ (\ref{eq:simple}),
must be modified for several reasons.  First of all, a correct
treatment of the perturbations produced during inflation would involve
a variational calculation \cite{mukhanov}. The results of this
formalism will be reported in another paper. The other reasons the
simple expression must be modified can be seen by recalling the
derivation of Eq.\ (\ref{eq:simple}).  Ignoring normalization for a
moment, the correct starting point is \cite{bst}
\begin{equation}
\label{eq:zeta}
\delta_H \propto \frac{\delta\rho}{\rho+p} \ .
\end{equation}
First consider the denominator of Eq.\ (\ref{eq:zeta}).  If only the
inflaton contributes to $\rho$ and $p$, then $\rho+p = \dot{\phi}^2$,
but around the resonance the fermions contribute to $\rho$ and $p$.
An expression for $\rho+p$ that includes this effect is $ \rho+p =
-\dot{H}m_{Pl}^2/4\pi$.

In the absence of fermion production, the numerator of Eq.\
(\ref{eq:zeta}) is just $\delta\rho = \delta\phi (dV/d\phi)$, with
$\delta\phi$ given by quantum fluctuations of the nonzero mode of the
inflaton field, $\delta\phi = H/2\pi$. However, fermions produced by
the resonance contribute to the fluctuations for two reasons:
fluctuations in the fermion field lead to fluctuations in the energy
density, and coupling of the produced fermions to the nonzero modes of
the inflaton field affect the perturbations.  Since the zero mode of
the inflaton field still dominates the energy density, these effects
are not expected to be large.  If they are included, they will
increase $\delta\rho$, hence increase the size of the feature in the
primordial power spectrum.  We will ignore the extra contributions to
$\delta\rho$ here, and simply assume that $\delta\rho =
\delta\phi(dV/d\phi) = (H/2\pi)(dV/d\phi)$.  Finally, one should not
use the slow-roll equation ($dV/d\phi=-3H\dot{\phi}$), because
$\ddot{\phi}$ may be large when fermions are produced.

In spite of the above considerations, we find that use of the simple
expression of Eq.\ (\ref{eq:simple}) allows reasonably
accurate description of the feature in the primordial power spectrum.
In the discussion below we describe both analytic and numerical
calculations using the following expressions for $\delta_H$:
\begin{equation}
\delta_H = \left\{ 
\begin{array}{ll}
\frac{\displaystyle{H^2}}{\displaystyle{5\pi\dot{\phi}}} & 
\qquad {\rm analytic} \\ 
&\\
\frac{\displaystyle{4H (dV/d\phi)}}{\displaystyle{15\dot{H}m_{Pl}^2}} 
& \qquad {\rm numerical} \ .
\end{array}
\right.
\end{equation}
For the analytic approximations we use $H^2=(8\pi/3)V/m_{Pl}^2$ and 
$\dot{\phi}$ determined by Eq.\ (\ref{eq:deltaphidot}).

\begin{figure}[t]
\centering \leavevmode\epsfxsize=350pt \epsfbox{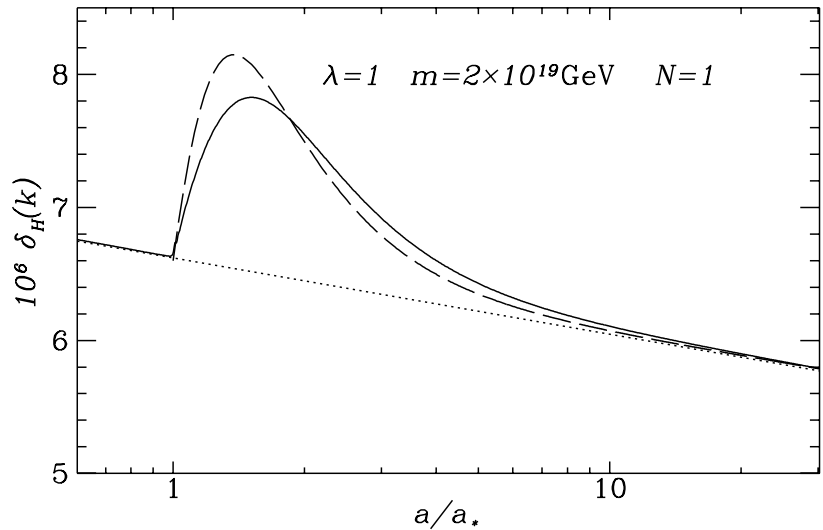}
\caption[fig4]{\label{delta_1_0} Resonant particle production produces
a peak in the perturbation spectrum as shown in the figure.  The solid
curve is the numerical result.  the dashed curve is the analytic form
of Eq.\ (\ref{eq:deltaanalytic}).  Finally, the dotted curve indicates
the power spectrum in the absence of resonant particle creation. }
\end{figure}

With the approximation that the change in $\dot{\phi}$ directly
translates into a change in the perturbation spectrum, the character
of the features produced in the spectrum can be understood from Eq.\
(\ref{eq:deltaphidot}).  Since $\Delta\dot{\phi}$ decays with a
characteristic width of $3H_*$, the width of the feature will
correspond to less than one e-fold of expansion.  For a given change
in $\dot{\phi}$ we can estimate the effect on the density
perturbation.  For the moment, assume the simple result for the
density perturbations of Eq.\ (\ref{eq:simple}).  For reasons
discussed above, we expect the width of the feature to be less than
one e-fold of expansion.  Defining $a_*$ to be the value of the scale
factor at resonance, a numerical calculation of the spectrum of
density perturbations is shown in Fig.\ \ref{delta_1_0} as a function
of $a/a_*$.  As expected, the width of the feature is less than a
single e-fold, and the peak amplitude of the feature corresponds to a
peak change in $\dot{\phi}$ estimated from Eq.\
(\ref{eq:estimatemaxphidot}).

An analytic expression for $\delta_H(k)$ can be obtained using the
analytic expression for $\dot{\phi}$.  The feature can be approximated
by a function of the form
\begin{equation}
\label{eq:deltaanalytic}
\delta_H(k) =  \frac{ [\delta_H(k)]_{\lambda=0}}
{1-\theta(a-a_*) |\dot{\phi}_*|^{-1}
N\lambda n_* H_*^{-1}(a_*/a)^3\ln(a/a_*)} \ .
\end{equation}
This expression is also shown in Fig.\ \ref{delta_1_0}.  Clearly it is
a very reasonable approximation.

In the comparison between analytic and numerical results presented in
this section we have assumed a single fermion species coupled to the
inflaton.  One might imaging a large number of fermion species coupled
to the inflaton.  In grand unified theories (GUTs) based upon gauge
groups such as $SO(10)$, $E_6$, and so on, representations of large
dimension are easily obtained.

If back reactions are ignored, then the effect of fermion production
on the perturbation spectrum is linear in $N$.  However, for large
values of $N\lambda^{5/2}$ the back reactions are important.  
We turn to such a case in the next section.

Before concluding this section we restate the important results:  The
feature in the spectrum should be about one e-fold in width, and the
amplitude of the feature should scale as $N\lambda^{5/2}$ until back
reactions become important.

Finally, we note that we can estimate how the change in the
perturbation spectrum scales with the inflaton potential.  Recalling
that $\delta_H=H_*^2/5\pi\dot{\phi}_*$, then
\begin{equation}
\label{eq:general}
\frac{\Delta \delta_H(k)}{\delta_H(k)} = 
-\frac{\Delta\dot{\phi}_*}{\dot{\phi}_*} = \frac{N\lambda^{5/2}}{6e\pi^3} \, 
\frac{3^{1/4}}{8^{3/4}} \, \frac{1}{\pi^{3/4}} \, 
\left( \frac{m_{Pl}^3(dV/d\phi)_*}{V_*^{3/2}}\right)^{1/2} \ ,
\end{equation}
where here we have used the slow-roll equation of motion,
$3H_*\dot{\phi}_* = -(dV/d\phi)_*$ and used $H_*^2=8\pi
V_*/3m_{Pl}^2$.  If we recall the approximation 
\begin{equation}
\delta_H \simeq \frac{8}{5}\, \left(\frac{8\pi}{2}\right)^{1/2} 
\left| \frac{V^{3/2}}{m_{Pl}^3dV/d\phi}\right| \ ,
\end{equation} 
we see that the term in parenthesis in Eq.\ (\ref{eq:general}) is
simply proportional to $\delta_H^{-1/2}$.  Therefore, if $dV/d\phi$
and $V$ are adjusted for a given $\delta_H$, then the relative
amplitude of the feature will be independent of $V$ and $dV/d\phi$.
This implies that the results of the paper can be applied to all
single-field inflation models, not just the large-field chaotic
inflation model we used.

\section{NUMERICAL RESULTS WHEN BACK REACTIONS ARE IMPORTANT}

We do not expect a qualitative difference in choosing $\phi_*\simeq
3m_{Pl}$, which corresponds to scales relevant for the CBR, and a lower
value of $\phi_*$.  Numerically it is convenient to work with smaller
$\phi_*$; in our runs we choose $\phi_*\sim 2m_{Pl}$.

\begin{figure}[t]
\centering \leavevmode\epsfxsize=350pt \epsfbox{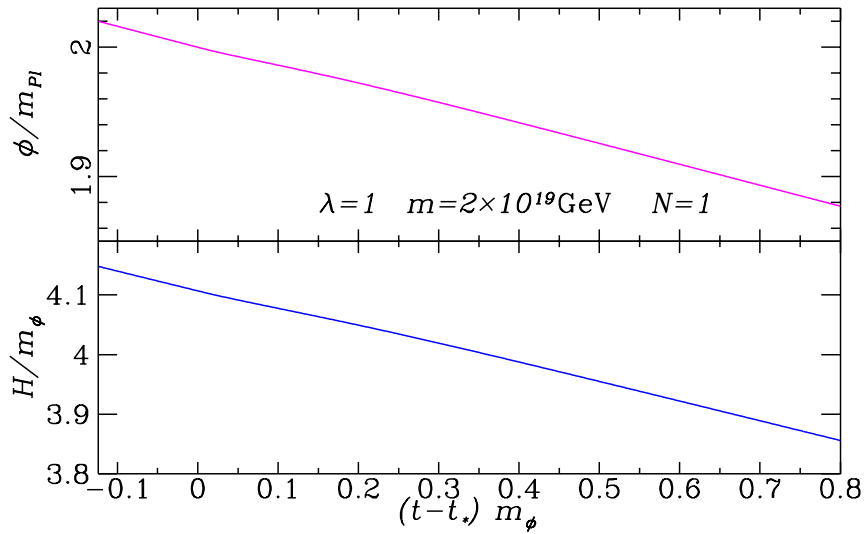}\\
\caption[fig5]{\label{h_phi_1_0} The evolution of $H$ and $\phi$ in
this model is not noticeably altered by resonant fermion production.
Only small inflections in $\phi(t)$ and $H(t)$ are noticeable around
the time of resonance, $t\simeq t_*$ where $t_*$ is the time when
$\phi=\phi_*$.}
\end{figure}

Since we already presented results for the case $\lambda=1$, $m =
2\times10^{19}$\,GeV, and $N=1$, for completeness we also present in
Fig.\ \ref{h_phi_1_0} the results for $\phi(t)$ and $H(t)$ in that
model.

Note that resonant fermion production has only a marginal effect on
the evolution of $\phi$ and $H$, but of course it has a large effect
on $\dot{\phi}$.  Near resonance, $\phi_* \simeq 2 m_{Pl}$ and $H_*
\simeq 4 m_\phi$.  The resulting feature in the perturbation spectrum
was shown in the previous section.

\begin{figure}[t]
\centering \leavevmode\epsfxsize=350pt \epsfbox{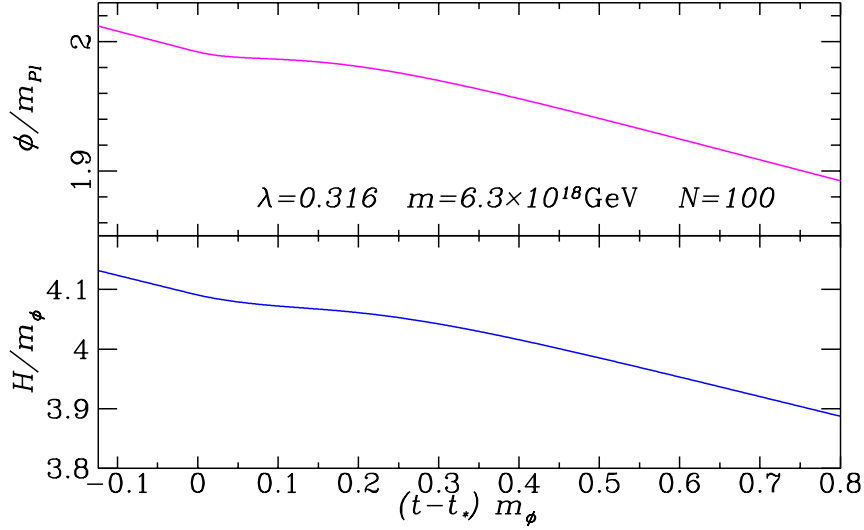}\\
\caption[fig6]{\label{h_phi_0_316} The evolution of $H$ and $\phi$ in
this model is noticeably altered by resonant fermion production.  The
flattening in the evolution of $\phi$ and $H$ begins at $t=t_*$ and
continues for a time of about $0.2m_\phi^{-1}$, after which $\phi$ and
$H$ resume decreasing at the rate they would if resonant particle
creation had not occurred.}
\end{figure}

Now let's study a model with an even larger amount of fermion
creation, where back reactions are quite important.  Let's consider
the case of $\lambda = 0.316$, $N=100$, and $m =
6.3\times10^{18}$\,GeV.  As in the previous example, the resonance
occurs for $\phi_*\simeq2m_{Pl}$.  The evolution of $\phi$ and $H$ in
this model is shown in Fig.\ \ref{h_phi_0_316}.  Clearly the evolution
of both $\phi$ and $H$ are modified by resonant particle production.
The change in $\dot{\phi}$ is even more dramatic than in the previous
example.

\begin{figure}[t]
\centering \leavevmode\epsfxsize=350pt \epsfbox{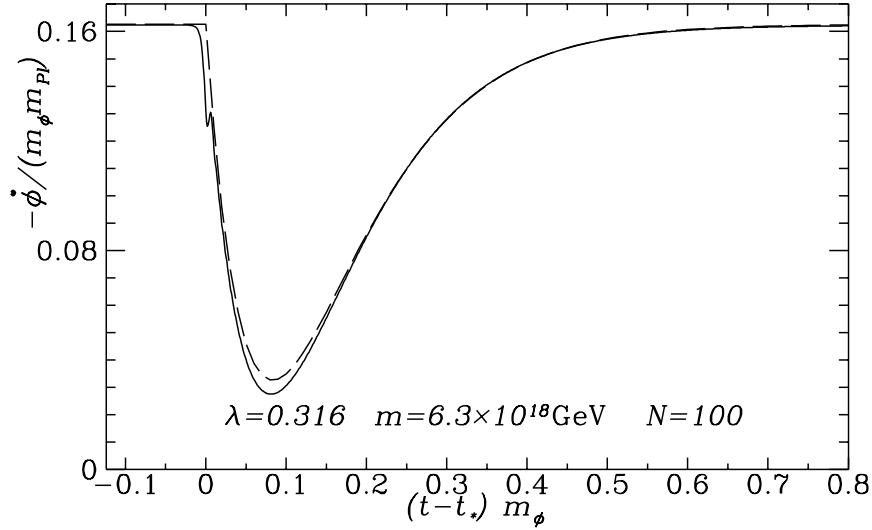}\\
\caption[fig7]{\label{phi_dot_0_316} The change in $\dot{\phi}$ in
this model is modified by the back reaction limiting the energy
extracted from the inflaton field.  Also shown by the dashed line is
the analytic expression of Eq.\ (\ref{eq:form}), with $A$ fit to agree
with ${\rm max}[\Delta\dot{\phi}]$.}
\end{figure}

The estimate of the change in $\dot{\phi}$ in the previous section
overestimates the true change.  The scaling of Eq.\
(\ref{eq:estimatemaxphidot}) predicts ${\rm
max}[\Delta\dot{\phi}]/m_{Pl}m_\phi$ of $100(0.316)^{5/2}$ times the
previous estimate of $0.03$.  This would predict a change in
$\dot{\phi}$ of $0.17$, which in fact is larger than $\dot{\phi}_*$.
The back reaction of the created fermions limits the value of ${\rm
max}[\Delta\dot{\phi}]$.  A numerical calculation of $\dot{\phi}$ in
this model is presented in Fig.\ \ref{phi_dot_0_316}.  While the
change in $\dot{\phi}$ is limited by the back reaction, nevertheless
it is quite large.

Even though the value of ${\rm max}[\Delta\dot{\phi}]$ is modified by the
back reactions, the feature still has the form 
\begin{equation}
\label{eq:form}
\Delta \dot{\phi}(t>t_*) = B(t-t_*)\exp[-3H_*(t-t_*)] \ , 
\end{equation}
where $B$ is a constant, $B< N\lambda n_*$.  The inequality is
saturated in the absence of back reactions.

\begin{figure}[t]
\centering \leavevmode\epsfxsize=350pt \epsfbox{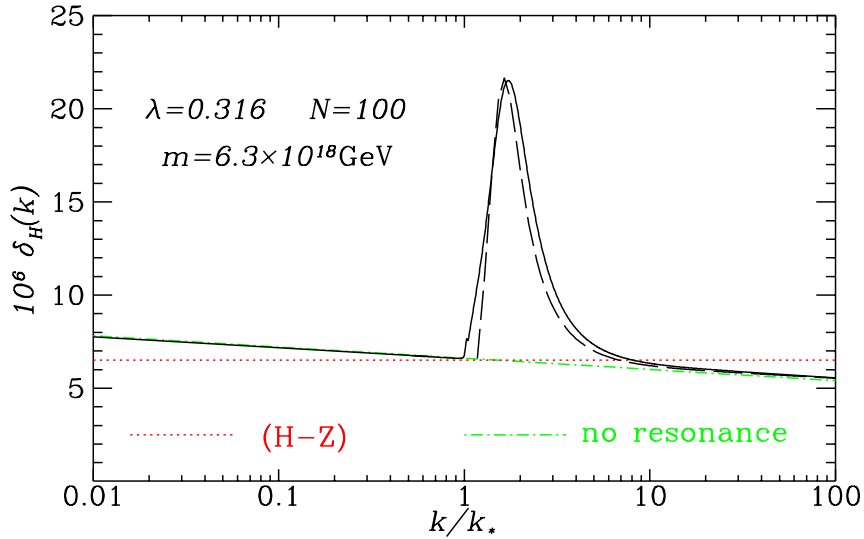}
\caption[fig8]{\label{delta_0_316} Resonant particle production
produces a peak in the perturbation spectrum as shown by the solid
curve in the figure.  Shown by a dashed curve is the analytic form of
$\delta_H(k)$ in Eq.\ (\ref{eq:fit}) with the coefficient $A$ chosen
to give the best fit.  Finally, also shown is a Harrison--Zel'dovich
spectrum and the spectrum in the inflation model without resonant
particle creation.}
\end{figure}

A numerical calculation of the primordial perturbation spectrum is
shown in Fig.\ \ref{delta_0_316} as a function of $k/k_*$.  A physical
wavenumber $k$ is related to $a$ by
\begin{equation}
\label{Ncross}
\ln \frac{k}{a_0 H_0} = 62 + \ln \frac{a}{a_*} + \ln \frac{a_*}{a_{\rm end}} 
        - \ln \frac{10^{16}\,{\rm GeV}}{V_k^{1/4}}
        + \ln \frac{V_k^{1/4}}{V_{{\rm end}}^{1/4}} - \frac{1}{3} \ln
        \frac{V_{{\rm end}}^{1/4}}{\rho_{{\rm reh}}^{1/4}} \ ,
\end{equation}
where the subscript `0' indicates present values; the subscript `$k$'
specifies the value when the wave number $k$ crosses the Hubble radius
during inflation ($k=aH$); the subscript `end' specifies the value at
the end of inflation; and $\rho_{{\rm reh}}$ is the energy density of
the universe after reheating to the standard hot big bang evolution.
The various values of $V$ denote the potential energy density at the
indicated epoch.  This calculation assumes that instantaneous
transitions occur between regimes, and that during reheating the
universe behaves as if matter-dominated.

Clearly the exact relationship between $a/a_*$ and a presently
observed value of wavenumber is model-dependent.  Although the
location of the peak in wavenumber (denoted by $k_*$), cannot be
unambiguously stated, the width of the feature in wavenumber is
proportional to the width of the feature in the ratio $a/a_*$, i.e.,
$k/k_*\propto a/a_*$. 

Also shown in Fig.\ \ref{delta_0_316} for comparison is a
Harrison--Zel'dovich (H--Z) spectrum (spectral index for the
primordial spectrum of $n=0$), and a numerical calculation of the
spectrum [assuming the spectrum is given by Eq.\ (\ref{eq:simple})] in
the model without resonant fermion production.  The effective spectral
index in the nonresonant inflation model is $n_S=1+d\ln\delta_H^2/d\ln
k = 0.96$.

This model produces a huge spike in the primordial power spectrum,
more than a factor of three times the underlying spectrum.  The feature
is fit by a function of the form of Eq.\ (\ref{eq:deltaanalytic}):
\begin{equation}
\label{eq:fit}
\delta_H(k) = \frac{\left[\delta_H(k)\right]_{\lambda=0}}
{1-\theta(k-k_*) 8.15A(k_*/k)^3\ln(k/k_*)} \ .
\end{equation}
If back reactions are ignored, then $ A = |\dot{\phi}_*|^{-1} N\lambda
n_* H_*^{-1}/8.15\simeq 0.23 N \lambda^{5/2}$.  The coefficient $A$ is
limited by back reactions to be such that the denominator is positive
for $k>k_*$; this implies $A\simlt 1$.  The first model ($N=1$,
$\lambda=1$) is unaffected by back reactions, and $A=0.23$.  Since
back reactions are important in the second model ($N=100$,
$\lambda=0.316$), the numerical fit ($A=0.7$) is smaller than
expected from the analytic expression $A =
0.23N\lambda^{5/2}=1.3$.

\section{DISCUSSION AND PROSPECTS FOR DETECTION OF FEATURES}

In this paper we have described a reasonably straightforward and
simple mechanism to produce features in the spectrum.  The features
can be characterized by a central location in wavenumber, a width, and
an amplitude.

The location of the peak in wavenumber depends on the value of $\phi_*
= m/\lambda$.  In the large-field inflation model we study,
$\phi_*$ has to be of order two to four times the Planck mass to have
any hope of being in the observationally accessible range.  Since to
produce a noticeable peak $\lambda$ cannot be too small, $m$ must
be of order $m_{Pl}$.

The narrow width of the feature, less than about a decade in
wavenumber (corresponding to about a single e-fold of inflation), is
reasonably independent of the model parameters.

The amplitude of the feature does depend on the model parameters, in
particular on the combination $N\lambda^{5/2}$.  In the first model
we presented ($\lambda=1$, $N=1$, $m=2\times10^{19}$\,GeV), the
amplitude of the feature was a factor of 1.3 larger than the
underlying spectrum.  The second model ($\lambda=0.316$, $N=100$,
$m=6.3\times10^{18}$\,GeV) had a peak increase in the power
spectrum of about 3.4 times the non-resonant result, or about 2.6
times the first model.  Scaling with $N\lambda^{5/2}$ would predict
a peak increase of the second model of 5.6 (rather than 2.6) times
that of the first model.  The failure of the peak amplitude to scale
as predicted is because of the back reaction limiting fermion
creation.

Our conclusion is that the characteristic signature of our mechanism
is a narrow spike in the power spectrum of width less than a decade in
wavenumber and amplitude anywhere from near zero to factors larger
than three.  An exceedingly good analytic form for the shape of the
feature is given in Eq.\ (\ref{eq:fit}).  The peak amplification is
$(1-A)^{-1}$ (recall that $A < 1$).  The first model had
$A=0.23$, while the second model had $A=0.7$.

We now turn to a discussion of how such a feature might be detected in
near-future experiments that will measure the present power
spectrum. Since the feature produced by resonant particle production
is approximately described by a single parameter ($0<A<1$), we can
estimate the sensitivity of large-scale structure surveys and CBR
experiments to a resonant particle production feature. The limit on
$A$ of course will depend on the value of $k_*$.

It is a relatively straightforward exercise to see how a feature in
the spectrum would appear in determination of the power spectrum from
large-scale structure surveys.  The first step is to convert the
primordial spectrum, $\delta_H(k)$, which is the amplitude of the
perturbation as wavenumber $k$ enters the horizon, to the present-day
power spectrum, $P(k)$, which describes the amplitude of the
perturbation at a fixed time.  This is done by means of the transfer
function, $T(k)$ \cite{efstathiou}:
\begin{equation}
\frac{k^3}{2\pi^2}P(k) = \left( \frac{k}{aH_0} \right)^4 T^2(k) 
\delta^2_H(k) \ .
\end{equation}
The transfer function depends upon a whole range of cosmological
parameters ($\Omega$, $H_0$, $\Lambda$, $\Omega_B$, and so on).  This
expression should be valid in the linear regime of the perturbations,
which in comoving wavenumber is approximately $k \simlt
0.2h$\,Mpc$^{-1}$.  Features in $\delta_H(k)$ are directly passed
through to $P(k)$, and in the linear regime should be straightforward
to identify.

For a volume-limited survey, the uncertainty due to cosmic variance
and shot noise in the estimated power per mode is
\cite{feldmankaiserpeacock,vogeley}
\begin{equation}
\label{eq:powererrors}
\frac{\Delta P(k)}{P(k)} = \sqrt{2\frac{(2\pi)^3}{V_SV_k}} 
\left[ 1+ \frac{1}{\bar{n}P(k)}\right] \ ,
\end{equation}
where $\bar{n}$ is the mean density of the sample of survey volume
$V_S$ and the estimates average over a shell in Fourier space of
volume $V_k=4\pi k^2\Delta k$.  The width of the bins is $\Delta k =
2\pi R^{-1}$, where $R$ is the depth of the survey.

Consider the sensitivity of determinations of the power spectrum
expected from the bright red galaxy (BRG) subsample of the SDSS.  The
volume-limited BRG subsample will include approximately $10^5$
galaxies in volume of effective depth $1h^{-1}$\,Gpc covering $\pi$
steradians.  It is expected that BRGs are biased by a factor of four
in the power spectrum.

\begin{figure}[t]
\centering \leavevmode\epsfxsize=350pt \epsfbox{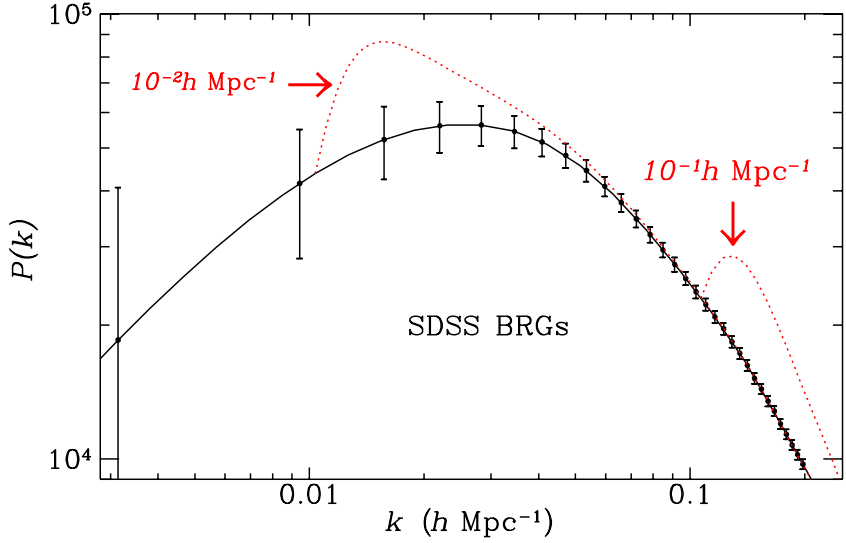}
\caption[fig9]{\label{bumps} This figure illustrates the predicted
uncertainties [see Eq.\ (\ref{eq:powererrors})] in the power spectrum
from the Bright Red Galaxy (BRG) subsample of the Sloan Digital Sky
Survey (SDSS).  The solid curve is a CDM power spectrum with $\Omega
h=0.25$.  The bins in $k$ and the errors are determined by Eq.\
(\ref{eq:powererrors}).  Also shown by the dotted curves are two
features in the power spectrum at the indicated values of $k_*$ and
for the same value of $A = 0.23$ [see Eq.\ (\ref{eq:fit})].}
\end{figure}

The sold curve in Fig.\ \ref{bumps} is generated assuming a
Harrison--Zel'dovich primordial spectrum and a CDM transfer function.
The points are in wavenumber bins with errors taken from Eq.\
(\ref{eq:powererrors}).  The two dotted curves are power spectra with
features of the form that would be generated by resonant particle
production.  In the form of Eq.\ (\ref{eq:fit}), the two examples have
$k_*=10^{-2}h$\,Mpc$^{-1}$ and $k_*=10^{-1}h$\,Mpc$^{-1}$ and both
have $A=0.23$ (the result of the model with $\lambda=1$, $N=1$, and
$m=2\times10^{19}$\,GeV).  Clearly either feature would be detected.

\begin{figure}[t]
\centering \leavevmode\epsfxsize=350pt \epsfbox{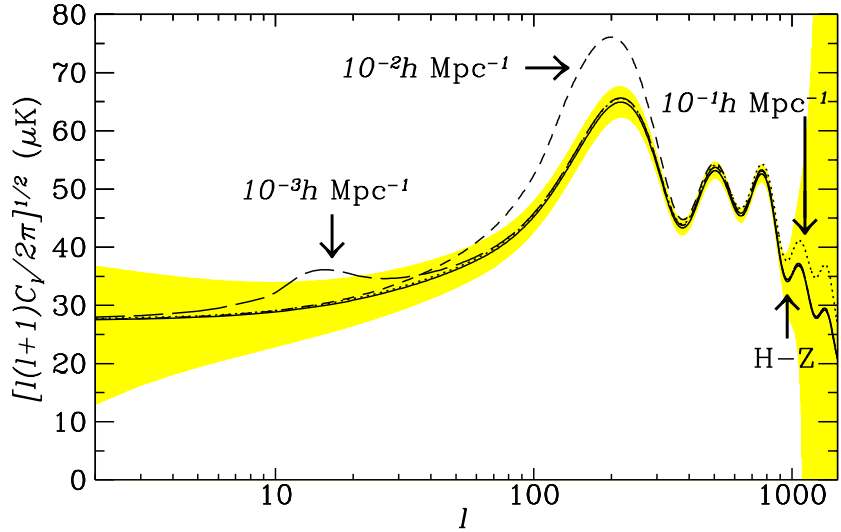}
\caption[fig10]{\label{cmb} The angular power spectrum calculated for a
featureless Harrison--Zel'dovich CDM model (solid curve marked H--Z)
and models with features produced by resonant particle production with
$A=0.23$ and the indicated values of $k_*$ [see Eq.\ (\ref{eq:fit})].
The cosmological parameters are identical for all models.  The shaded
area indicates the expected uncertainty in the $C_l$'s from the MAP
satellite. The models have all been normalized to the same value of
$C_2$.}
\end{figure}

Features in the primordial power spectrum will also affect the CBR
temperature fluctuations.  The CBR temperature fluctuations can be
expanded in spherical harmonics, $ \delta T/T =\sum_l\sum_m
a_{lm}Y_{lm}(\theta,\phi)$ ($2<l<\infty$ and $-l<m<l$). The
anisotropies are described by the angular power spectrum, $C_l=
\langle |a_{lm}|^2\rangle$, as a function of multipole number $l$.

The amplitude of the perturbation spectrum at a given wavenumber $k$
contributes to the angular power spectrum in a range of $l$, so the
effect of a narrow feature is spread over several values of $l$.  The
angular power spectrum is easily calculated using CMBFAST
\cite{cmbfast}.  The result of a sample model calculations is shown in
Fig.\ \ref{cmb}.  The angular power spectrum resulting from a
featureless power-law spectrum (Harrison--Zel'dovich) is shown by the
solid curve.  Also shown for comparison are three other models with
features in the primordial power spectrum of the form in Eq.\
(\ref{eq:fit}), with $A = 0.23$ and $k_* = 10^{-3}h$\,Mpc$^{-1}$,
$10^{-2}h$\,Mpc$^{-1}$, and $10^{-1}h$\,Mpc$^{-1}$.

The CBR is sensitive to the power spectrum in the interval from about
$10^{-4}h$\,Mpc$^{-1}$ to $10^{-1}h$\,Mpc$^{-1}$.  For
$k_*<10^{-3}h$\,Mpc$^{-1}$, a feature with $A<0.23$ will be too small
to be detected because cosmic variance will be large at small $l$.
For $k_*\sim10^{-2}h$\,Mpc$^{-1}$, the largest effect will be near the
first Zel'dovich peak and could have an effect on determination of the
geometry of the universe from the location of the peak.  If
$k_*\sim10^{-1}h$\,Mpc$^{-1}$, the largest effect will be in the
region of the secondary peaks.  If $k_*$ is much larger than
$10^{-1}h$\,Mpc$^{-1}$, the effect will be out of the range of CBR
sensitivity.

\begin{figure}[t]
\centering \leavevmode\epsfxsize=350pt \epsfbox{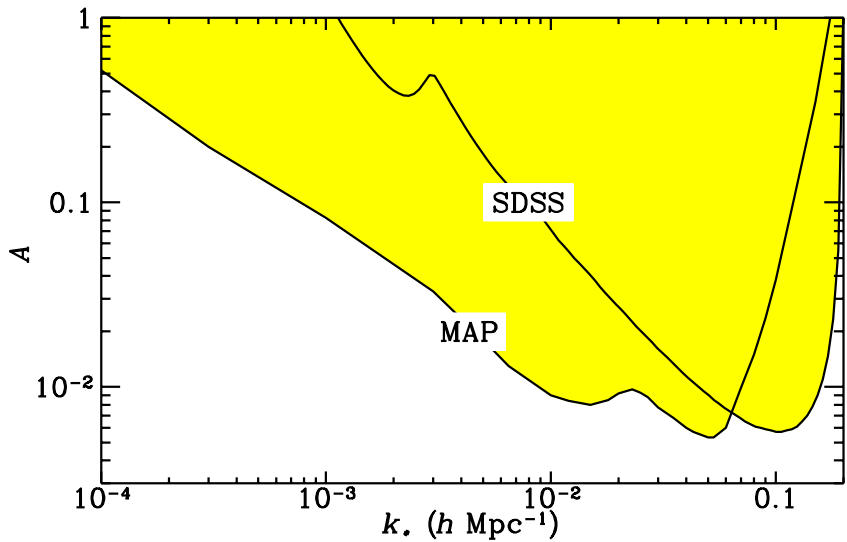}
\caption[fig11]{\label{chisqd} The shaded region will produce a
$2\sigma$ change in the matter power spectrum with the expected
sensitivity expected for the SDSS--BRG subsample and the MAP CBR
satellite mission.}
\end{figure}

A meaningful detection limit is hard to quote; the limiting value of
$A$ depends on $k_*$ (among other things).  In Fig.\ \ref{chisqd} we
indicate by the shaded regions the values of $A$ that will differ
more than a $2\sigma$ from the best-fit featureless power-law spectrum 
expected to be determined by SDSS and MAP.

The feature around $k_*=3\times 10^{-3}h$\,Mpc$^{-1}$ in the SDSS
limit is due to the feature falling between the widely spaced bins in
that region.  In the figure we have assumed the normalization will be
set by CBR determinations, i.e., the normalization is a free parameter
for the MAP limits, but not for the SDSS limits.  In principle, the region of
sensitivity could extend to $k_*$ greater than $0.1h$\,Mpc$^{-1}$ if the
evolution of the power spectrum in the mildly nonlinear regime was
completly understood.

The form of the limit from MAP around $k_* \sim 3\times10^{-2}h$\,Mpc$^{-1}$
is due to the fact that the spectral feature in that region falls in the region of
the Zel'dovich peaks.

For $ 10^{-2} \simlt k_*\simlt 10^{-1}h$\,Mpc$^{-1}$, the smallest $A$
that has a $2\sigma$ effect is about $6\times10^{-3}$.\footnote{Here
we are not saying that $A=6\times10^{-3}$ would be cleanly detectable,
only that a featureless power-law spectrum would not be a good fit.}
This would correspond to the limit
\begin{equation}
\lambda < 0.2/N^{2/5}  \  .
\end{equation}

The correct way to interpret this limit is that if it is violated, the
feature in the spectrum produced by resonant particle production has
the potential of being detected in large-scale structure surveys.

Again, any limit on $A$ will depend on $k_*$.  For $k_*$ near
$10^{-2}h$\,Mpc$^{-1}$ and $A$ as small as 0.01, the effect of
resonant particle production will be noticeable.  For larger
$A$ it is possible that the feature could be noticeable in both
CBR and large-scale structure surveys.

\section{CONCLUSIONS}

For the most part, analysis involving the power spectrum assumes that
it is a featureless power law.  Occasionally one includes smooth
variation in the spectral index, but usually retaining the assumption
that the power spectrum is featureless.  

In this paper we presented a reasonably straightforward scenario for
producing a spike (or several spikes) in the power spectrum.  While
analysis assuming a featureless power spectrum is certainly a
reasonable first step, it would be wise to keep in mind that when
doing so one is making an approximation about the dynamics during
inflation that may not be true.

Let us now elaborate about the particle physics motivations of our
work. When speaking about a `model of inflation', the phrase is
actually used by the community in two rather different ways.  At the
simplest level a `model of inflation' is taken to mean a form for the
potential, as a function of the fields giving a significant
contribution to it.  In single-field models this is just the inflaton
field $\phi$. If one knows $V(\phi)$ and the field value at the end of
inflation, this allows one to calculate the spectrum of density
perturbations.  

At a deeper level, one thinks of a `model of inflation' as something
analogous to the Standard Model of particle interactions \cite{lyth}.
One imagines that nature has chosen some extension of the Standard
Model, and that the scalar fields relevant for inflation are part of
that model. In this sense a `model of inflation' is more than merely a
specification of the the potential of the relevant fields. It will
provide answers to some fundamental questions such as whether the
relevant fields and interactions have already been invoked in some
other context and, if so, what are the mass parameters and the
coupling constants in the sectors coupled to the inflaton field(s).

Of course, it would have been wonderful if inflation already dropped
out of the Standard Model of particle physics, but sadly that is not
the case.  On the contrary, it seems very likely that the physics
responsible for an inflationary stage during the early evolution of
the universe is due to some particle physics model whose typical
energy is much larger than the weak scale and whose field values
during inflation may be larger than the Planck mass.

Well studied extensions of the Standard Model, such as supergravity
and superstring theories, have the Planck mass as a fundamental
scale. The exact motivations and goals of these theories beyond the
Standard Model might be different, but for many applications to
cosmology they have several common features. For instance, they
contain in the spectrum particles with Planckian masses and are
formulated in extra-dimensions. These extra-dimensions are compact and
smaller than the three large spatial dimensions. It is therefore
possible to dimensionally reduce the system and obtain and `effective'
$3+1$ dimensional theory leaving behind a tower of Kaluza-Klein (KK)
states (pyrgons \cite{ks}) whose mass is of the order of the inverse
size of the extra dimension. Since the extra dimensions are expected
to have a size characteristic of the Planck length, these KK states
therefore have Planckian masses. Furthermore, these states are largely
degenerate, the level of degeneracy depending upon the geometrical
structure of the compact space.  It is therefore quite natural to
expect that a large number of fermions with the same mass,
approximately $ m_{Pl}$, couple to the sector responsible for the
inflationary stage. In the simplest approach, this degeneracy factor
may be accounted for by the parameter $N$ we have used in the previous
sections. Since $N$ may be easily of the order of 100, a detectable
signature in the spectrum may be present even in the case of small
Yukawa couplings.  Thus, the observation of a spike (or several
spikes) in the power spectrum of the density perturbation may help us
in understanding the features of the inflaton sector and thus testing
Planckian physics. In the absence of any other experimental signature,
this is certainly very intriguing.

\acknowledgements{DJHC was supported by the Department of Energy.  
EWK was supported by the Department of Energy and NASA under Grant
NAG5-7092.  We benefitted from the advice and encouragement of Alexi
Starobinski, as well as conversations with Andrew Liddle and Dan
Eisenstein.  EWK would like to acknowledge the hospitality
of the Isaac Newton Institute of Mathematical Sciences in Cambridge
and the Heisenberg Institute in Munich for their hospitality during
the course of the work presented here.  IIT would like to thank the Isaac
Newton Institute of Mathematical Sciences in Cambridge for their hospitality.}


\end{document}